\documentclass{kapproc} % Computer Modern font calls
\usepackage{epsfig}
\def\lesssim{\mathrel{\hbox{\rlap{\hbox{\lower4pt\hbox{$\sim$}}}\hbox{$<$}}}}
\def\gtrsim{\mathrel{\hbox{\rlap{\hbox{\lower4pt\hbox{$\sim$}}}\hbox{$>$}}}}

\let\footnote\savefootnote

\let\footnotetext\savefootnotetext 

\kluwerbib
\begin{document}

\articletitle[Stellar collapse and gravitational waves]{Stellar
collapse and\\ gravitational waves}

\author{Chris L.\ Fryer} \affil{Theoretical Astrophysics, Los Alamos
National Laboratories, Los Alamos, NM, 87545}

\author{Daniel E.\ Holz and Scott A.\ Hughes} \affil{Kavli Institute for
Theoretical Physics, UC Santa Barbara, Santa Barbara, CA 93106}

\author{Michael S. Warren} \affil{Theoretical Astrophysics, Los Alamos
National Laboratories, Los Alamos, NM, 87545}

\begin{abstract}

The new generation of gravitational wave (GW) detectors have the
potential to open a novel window onto the violent dynamics of core
collapse.  Although it is certain that core collapse events generate
gravitational radiation, understanding the characteristics of the
radiation --- whether it can be measured with these detectors, and the
best way to go about doing so --- is a challenging problem. In this
chapter we review the promise of GWs as observational probes,
including a discussion of the current state of GW detectors, and
discuss the status of work to understand the waves generated by
stellar core collapse.

\end{abstract}

\section{Introduction}
\label{sec:intro}

For quite some time the collapse of massive stellar cores in supernova
explosions has been regarded as likely to be an interesting and
important source of gravitational radiation (see, for example,
{\cite{de1983}} for an early review of the subject).  Core collapse
events certainly produce GWs: large amounts of mass ($\sim
1$--$100\,M_\odot$) flow in a compact region ($\sim 10^8$--$10^9$ cm)
at relativistic velocities ($v/c\sim 1/5$).  These characteristics are
{\it necessary} conditions for a source to be an interesting GW
source; however, they are {\it not} sufficient conditions.  Detailed
analysis is needed to understand whether these potentially interesting
sources are in fact sufficiently asymmetric to be strong radiators,
and whether we can understand their robust features well enough to
search for them with GW observatories.

Theoretical modeling of stellar collapse has become
increasingly sophisticated in recent years: collapse
theorists can now model collapses in three dimensions
(albeit requiring enormous amounts of CPU time), and can
include important physics such as neutrino transport and the
equation of state of very dense matter.  Previously, most
estimates of GW emission from stellar collapse were (of
necessity) rather idealized, typically studying the
evolution of instabilities in rotating fluid distributions
(\cite{chandra1969}, \cite{cent_etal2000}, \cite{new2000}).
Although extremely valuable in establishing the conditions
under which strongly radiating instabilities can occur,
these analyses do not incorporate important physical effects
which occur during stellar collapse.  Using modern collapse
models, estimates of GW emission can now be constructed that
more realistically reflect the phenomenology of stellar
collapse.  Coupling these modern models with the wisdom that
has been gleaned from decades of study from the idealized
cases, we should now be able to assemble a robust picture of
the GWs arising from core collapse.

The maturity of modern collapse simulations is quite timely:
they arrive just as broad-band GW detectors based on laser
interferometry are coming into existence.  Science runs with
the initial generations of these detectors have just begun,
although these instruments have not yet reached their design
sensitivities. Furthermore, the design sensitivities are not
really at the level where stellar core collapse can be
considered a realistically interesting signal (\cite{fhh},
hereafter FHH).  However, the detectors' sensitivites have
improved quite rapidly\footnote{The LIGO detectors have
improved their sensitivities by several orders of magnitude
across a wide frequency band in the first nine months of
2002.}, and a vigorous R\&D program promises to push the
sensitivites to better levels very rapidly
(\cite{whitepaper}).  The possibility of seeing core
collapse GWs is stronger than ever\footnote{It's worth
noting that there is always a chance---albeit a small
one---that a relatively nearby star could go supernova and
produce an anomalously strong signal. The current
configuration of LIGO would likely have seen SN 1987a, were
it to have been operating back then. The last thing that the
GW detection community wants is to miss another such event!}.

In this chapter we review the state of our present knowledge
of GW events.  We first present an overview of GW physics,
focusing on the properties of the waves themselves and how
one detects them, plus a brief review of the current state
of the various GW interferometers.  (We note that the
discussion of the detectors' state is quite likely to be out
of date by the time this article appears in print.)  This
material is largely taken from a recent review article
(\cite{snowmass}) with some updating to reflect the current
status of the detectors.  We then turn to an
in-depth discussion of GWs from core collapse events,
comparing the likely properties of these waves to the
detection thresholds of the new interferometric detectors.
This discussion is based to a large extent on the review of
FHH, with some updating based on recent work on r-modes and
$\mbox{3-D}$ simulations.

\section{GWs and detectors: overview}
\label{sec:overview}

Gravitational radiation is a natural consequence of general
relativity, first described more-or-less correctly by Albert Einstein
(1918).  GWs are tensor perturbations to the metric of spacetime,
propagating at the speed of light, with two independent polarizations.
As electromagnetic radiation is generated by the acceleration of
charges, gravitational radiation arises from the acceleration of
masses.  Electromagnetic waves are created (at lowest
order) by the time changing charge dipole moment, and are thus dipole
waves.  Monopole EM radiation would violate charge conservation.  At
lowest order, GWs come from the time changing quadrupolar distribution
of mass and energy; monopole GWs would violate mass-energy
conservation, and dipole waves violate momentum conservation.

\begin{figure}[t]
\includegraphics[width = 12cm]{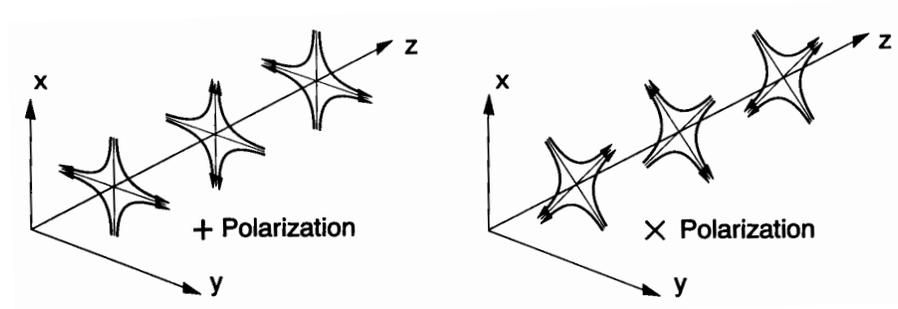}
\caption{The lines of force associated with the two polarizations of a
GW (from {\cite{ligoscience}}).}
\label{fig:forcelines}
\end{figure}

GWs act tidally, stretching and squeezing objects as they pass
through.  Because the waves arise from quadrupolar oscillations, they
are themselves quadrupolar in character, squeezing along one axis
while stretching along the other.  When the size of the object that
the wave acts upon is small compared to the wavelength (which is the
case for all ground-based detectors), forces arising from the two GW
polarizations act as in Fig.\ {\ref{fig:forcelines}}.  The
polarizations are named ``$+$'' (plus) and ``$\times$'' (cross),
as a result of the orientation of the axes associated with their force
lines.

\begin{figure}[t]
\includegraphics[width = 12cm]{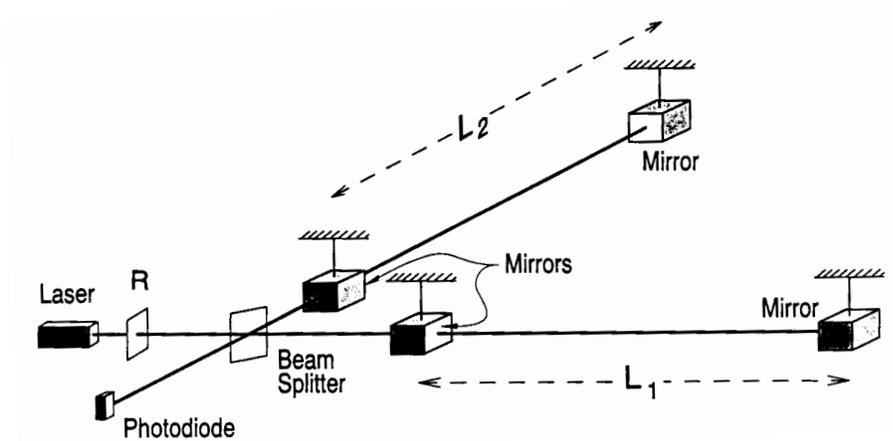}
\caption{Layout of an interferometer for detecting GWs
(from {\cite{ligoscience}}).}
\label{fig:ifo}
\end{figure}

Interferometric GW detectors measure this tidal field via their action
upon a widely-separated set of test masses, arranged as in Fig.\
{\ref{fig:ifo}}.  Each mass is suspended with a sophisticated pendular
isolation system to eliminate the effect of local ground noise.  Above
the resonant frequency of the pendulum (typically of order $1\,{\rm
Hz}$), the mass moves freely.  In the absence of a GW, the sides $L_1$
and $L_2$ shown in Fig.\ {\ref{fig:ifo}} are taken to have
the same length, $L$.

Suppose the interferometer in Fig.\ {\ref{fig:ifo}} is
arranged so that its arms lie along the $x$ and $y$ axes of
Fig.\ {\ref{fig:forcelines}}.  Suppose further that a wave
impinges on the detector down the $z$ axis, and the ``$+$''
polarization axes are aligned with the detector's arms.  The
tidal force of this wave stretches one arm while squeezing
the other; each arm oscillates between stretch and squeeze
as the wave itself oscillates.  The wave is detectable by
measuring the separation of the test masses in each arm.  In
particular, since one arm is always stretched while the
other is squeezed, we can monitor the length difference of
the two arms:
\begin{equation}
\delta L(t) \equiv L_1(t) - L_2(t)\;.
\label{eq:delta_L}
\end{equation}
For the case discussed above, this change in length turns out to be
the armlength times the $+$ polarization amplitude:
\begin{equation}
\delta L(t) = h_+(t)L\;.
\label{eq:h_simple}
\end{equation}
The GW acts as a dimensionless strain in the detector; $h$ is often
referred to as the ``wavestrain''.  Equation (\ref{eq:h_simple}) is
derived by applying the equation of geodesic deviation to the
separation of the test masses, using a GW tensor on a flat background
spacetime to develop the curvature tensor; see Thorne (1987), Sec.\
9.2.2 for details.  We obviously do not expect astrophysical GW
sources to align themselves in as convenient a manner as described
above.  Generally, both polarizations of the wave influence the test
masses:
\begin{equation}
{\delta L(t)\over L} = F^+ h_+(t) + F^\times h_\times(t) \equiv h(t)\;.
\label{eq:h_def}
\end{equation}
The antenna response functions $F^+$ and $F^\times$ weight the two
polarizations in a quadrupolar manner as a function of a source's
position and orientation relative to the detector; see Thorne (1987),
Eqs.\ (104a,b) and associated text.

The test masses at the ends of each arm are made of a highly
transparent material (fused silica in present designs; perhaps
sapphire in future upgrades). All of the test masses are faced with
dielectric coatings such that they are extremely good reflectors of
the 1.064 micron infrared laser light. The mirrors at the far end of
each arm have amplitude reflectivities approaching unity.  The mirrors
at the corner joining the arms are less reflective, since they must
couple the light into the Fabry-Perot cavity arms.  The corner
mirrors' multilayer dielectric coatings have power reflectivities $T
\sim 3\%$.  A very stable laser beam is divided at the beamsplitter,
directing light into the two arm cavities.  If the finesse of the
cavity is ${\cal F}$ and the amplitude reflectivity of the corner
mirrors is $r_{\rm corner}$, then each photon makes on average ${\cal
F}/\pi\simeq\sqrt{r_{\rm corner}}/(1 - r_{\rm corner}) \sim 65$
bounces.  The light from the two arms then recombines at the
beamsplitter.  The mirrors are positioned so that, in the absence of a
GW, all of the light goes back to the laser and the photodiode reads
no signal.  If a signal is present, the relative phase $\Phi$ of the
two beams must have changed by an amount proportional to $h$, changing
the light's interference pattern.  With no intervention, this would
cause light to leak into the photodiode.  In principle, the wavestrain
$h$ could be read from the intensity of this light.  In practice, a
system of servo loops controls the system such that destructive
interference is guaranteed --- the photodiode is kept dark, and is
thus called the ``dark port''.  The wavestrain $h$ is encoded in the
servo signals used to keep the dark port dark.

The energy flux carried by GWs scales with $\dot h^2$ (where
the overdot represents a time derivative). To conserve
energy flowing through large spheres, the wavestrain falls
off with distance as $1/r$. We have already argued that the
lowest order contribution to the waves is due to the
changing quadrupole moment of the source.  To order of
magnitude, this moment is given by $Q \sim (\mbox{source
mass})(\mbox{source size})^2$.  Since the wavestrain is
dimensionless, the scaling must take the form
\begin{equation}
h \sim {G\over c^4}{\ddot Q\over r}\;.
\label{eq:h_ordermag}
\end{equation}
The second time derivative of the quadrupole moment is given
approximately by $\ddot Q\simeq 2M v^2\simeq 4 E^{\rm ns}_{\rm kin}$,
where $v$ is the source's internal velocity, and $E^{\rm ns}_{\rm
kin}$ is the nonspherical part of its internal kinetic energy.  Strong
sources of gravitational radiation are sources that have strong
non-spherical dynamics; hence, core collapse events must be quite
asymmetric in order to give off considerable energy in GWs.

For an interesting rate of observable events, detectors must be
sensitive to sources at rather large distances.  For example, to
detect several stellar collapse event in a year, our detectors must
reach $r \sim 10$ Mpc.  For stellar collapse, a reasonable
estimate of the non-spherical kinetic energy is $E^{\rm ns}_{\rm
kin}/c^2\sim 0.1\,M_\odot$.  Plugging these numbers into Eq.\
(\ref{eq:h_ordermag}) yields the estimate
\begin{equation}
h \sim 10^{-21}\mbox{--}10^{-22}\;.
\label{eq:h_num_est}
\end{equation}
These tiny numbers set the sensitivity required to measure
GWs.  Combining this
scale with Eq.\ (\ref{eq:h_def}) says that for every kilometer of
baseline $L$, we must measure a distance shift between the
arm lengths, $\delta L$, of better
than $10^{-16}$ centimeters.

The prospect of achieving such stringent displacement sensitivities
often strikes people as insane.  How can light, whose wavelength
$\lambda\sim 10^{-4}\,{\rm cm}$ is $10^{12}$ times larger than the
typical displacement, be used to measure that displacement?  For that
matter, how is it possible that thermal motions don't wash out these
tiny effects?

That such measurement is possible with laser interferometry was first
analyzed thoroughly by Weiss (1972).  \footnote{It should be
noted that the possibility of detecting GWs with laser interferometers
has an even longer history, reaching back to Pirani in 1956,
and was independently proposed by several workers: Gertsenshtein and
Pustovoit in 1962, Weber in the 1960s, and Weiss c.\ 1970.  See Sec.\
9.5.3 of Thorne (1987) for further discussion and references.}
We examine first how a 1 micron laser can measure a $10^{-16}$ cm
effect.  As mentioned above, the light bounces back and
forth roughly 100 times
before leaving the arm cavity (corresponding to about half a cycle of
a 100 Hz GW).  The light's accumulated phase shift during those 100
round trips is
\begin{equation}
\Delta\Phi_{\rm GW}\sim100\times2\times\Delta L\times2\pi/\lambda
\sim 10^{-9}\;.
\label{eq:phase_estimate}
\end{equation}
This phase shift is measurable provided that the shot noise at the
photodiode, $\Delta\Phi_{\rm shot}\sim 1/\sqrt{N}$, is less than
$\Delta\Phi_{\rm GW}$.  $N$ is the number of photons accumulated over
the measurement; $1/\sqrt{N}$ is the magnitude of phase fluctuation in
a coherent state, appropriate for describing a laser.  We therefore
must accumulate $10^{18}$ photons over the roughly $0.01$ second
measurement, which translates to a laser power of about 100 watts.  In
fact, as was pointed out by Drever (1983), one can use a much less
powerful laser: even in the presence of a GW, only a tiny portion of
the light that comes out of the interferometer's arms goes to the
photodiode.  The vast majority of the light is sent back to the laser.
An appropriately placed mirror bounces this light back into the arms,
{\it recycling} it.  The recycling mirror is shown in Fig.\
{\ref{fig:ifo}}, labeled ``R''.  With that mirror, a laser of $\sim
10$ watts drives several hundred watts to circulate in the ``recycling
cavity'' (the optical cavity between the recycling mirror and the
arms), and $\sim 10$ kilowatts to circulate in the arms.

Thermal excitations are overcome by averaging over many many
vibrations.  For example, the atoms on the surface of the
interferometers' test mass mirrors oscillate with an amplitude
\begin{equation}
\delta l_{\rm atom} = \sqrt{k T\over m\omega^2}
\sim 10^{-10}\,{\rm cm}
\end{equation}
at room temperature $T$, with $m$ the atomic mass, and with a
vibrational frequency $\omega\sim10^{14}\,{\rm s}^{-1}$.  This
amplitude is huge relative to the effect of the GW---why doesn't it wash
out the wave?  GWs are detectable because the atomic
vibrations are random and incoherent.  The $\sim 7$ cm wide laser beam
averages over about $10^{17}$ atoms and at least $10^{11}$ vibrations
in a typical measurement.  These atomic vibrations cancel
out, becoming irrelevant compared
to the coherent effect of a GW.  Other thermal vibrations, however,
end up dominating the detectors' noise spectra
in certain frequency bands.  For example, the test masses' normal
modes are thermally excited.  The typical frequency of these modes is
$\omega\sim 10^5\,{\rm s}^{-1}$, and they have mass $m \sim 10\,{\rm
kg}$, so $\delta l_{\rm mass} \sim 10^{-14}\,{\rm cm}$.  This, again,
is much larger than the effect we wish to observe.  However, the modes
are very high frequency, and so can be averaged away provided the test
mass is made from material with a very high quality factor $Q$ (so
that the mode's energy is confined to a very narrow band near
$\omega$, and thus doesn't leak into the band we want to use for
measurements).  Understanding the physical nature of noise in GW
detectors is an active field of current research; see Levin (1998),
Liu \& Thorne (2000), Santamore \& Levin (2001), Buonanno \& Chen
(2001a,b), Hughes \& Thorne (1998), Creighton (2000), and references
therein for a glimpse of recent work.  In all cases, the fundamental
fact to keep in mind is that a GW acts {\it coherently}, whereas noise
acts {\it incoherently}, and thus can be beaten down provided one is able
to average away the incoherent noise sources.

\subsection{Current detectors}
\label{sec:groundbased}

The first generation of long baseline, kilometer-scale interferometric
GW detectors are in operation or are being constructed and
commissioned at several sites around the world.  Briefly, the major
ground-based interferometric GW projects are as follows:

\begin{itemize}
\item \textbf{LIGO}.  Three LIGO (Laser Interferometer
Gravitational-wave Observatory) interferometers are currently
operating: two in Hanford, Washington (with 2 and 4 km arms, sharing
the same vacuum system), and one in Livingston, Louisiana (4 km arms).  An
aerial view of the Hanford site is included in Fig.\
{\ref{fig:ligo_optical}}.  The LIGO detectors are designed to operate
as power recycled Michelson interferometers with arms acting as
Fabry-Perot cavities.  The large distance between sites (about $3000$
km) and differing arm lengths are designed to support coincidence
analysis.  Much current research and development is focused on
advanced LIGO detector design.  The goal of these planned improvements
is to provide a broader frequency band and a $\sim$ 10-fold increase
in range for sources via a lowered noise floor.

\item \textbf{Virgo}.  Virgo, the Italian/French long baseline GW
detector, is under construction near Pisa, Italy (\cite{marion2000}).
It has 3 km arms and advanced passive seismic isolation systems.  In
most respects, Virgo is similar to LIGO; a major difference is
that it should achieve better low frequency sensitivity in
its first generation due to its advanced seismic isolation.  Virgo will
usefully complement the LIGO detectors, strengthening coincidence
analysis and significantly improving source position determination.

\item \textbf{GEO600}.  GEO600 is a 600 meter interferometer
constructed by a British-German collaboration near Hannover, Germany
(\cite{luck2000}).  It uses advanced interferometry and advanced
low noise multiple pendulum suspensions, serving as a testbed for
advanced detector technology, and allowing it to achieve sensitivities
comparable to the multi-kilometer instruments.

\item \textbf{TAMA300}.  The TAMA detector near Tokyo, Japan can
already claim significant observation time, with more than 1000 hours
of operation (\cite{andoetal}).  It has achieved a peak strain
sensitivity of $h\sim 3\times 10^{-19}\,{\rm Hz}^{-1/2}$ at
frequencies near 1000 Hz.  TAMA has 300 meter arms and is operated in
the recombined Michelson configuration with Fabry-Perot arms.  A much
improved 3 km detector is currently under design (\cite{kuroda2000}).

\item \textbf{ACIGA}.  The Australian Consortium for Interferometric
Gravitational Astronomy plans to build an observatory near Perth,
Australia (\cite{mccleland2000}).  They are presently engaged in the
construction of an 80 meter research interferometer, which can be
extended to kilometer scale.  They are studying advanced detection
methods and technologies which could lead to much decreased noise
floors in advanced interferometers.

\end{itemize}

Since interferometric GW detectors have nearly equal sensitivity in
all directions, it is essentially impossible to deduce pointing
information from the output of a single detector.  To get accurate
information about the source direction it is necessary to make use of
time-of-flight differences between detectors.  To do this the detectors must
be widely spaced, and cannot be collinear.  At the very minimum, three sites
are needed for acceptable pointing.  A fourth detector, widely removed
from the plane of the other three, is particularly valuable for
improving directional information.  Thus a detector in Australia would
greatly add to the science output of the GW observatory network, which
is otherwise confined entirely to the northern hemisphere.

LIGO is representative of the design and operation principles of these
interferometers, so we shall focus on it for the remainder of our
discussion.

\subsection{LIGO overview}

Construction of both LIGO facilities and vacuum systems was completed
in early 2000.  Both observatories have mostly concentrated to date on
detector commissioning and a series of ``Engineering Runs'',
leading to the first ``Science Run'', which began in August 2002.  This will
be followed by more frequent and longer Science Runs until 2006/7,
when major detector upgrades are scheduled.  In parallel with the
commissioning effort and Engineering/Science Runs, the LIGO collaboration
is also focused on the research and development of advanced
detectors, promising a wider detector band and greatly improved
sensitivity.

\begin{figure}[t]
\includegraphics[width = 12cm]{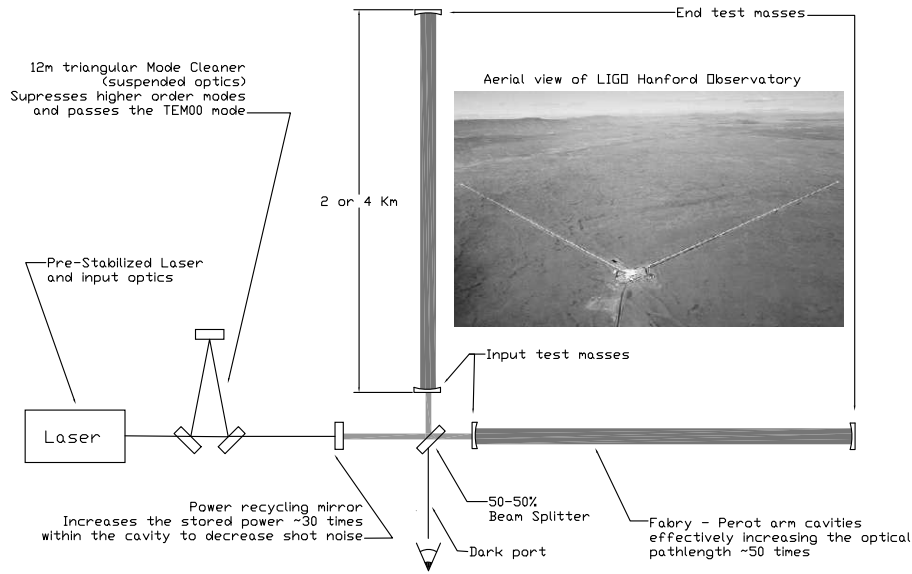}
\caption{Simplified optical layout of a LIGO interferometer.  Shown
here are the prestabilized laser, the input mode cleaner, the
recycling mirror, and the test mass mirrors.  As discussed in Sec.\
{\ref{sec:overview}}, servo loops ensure that the recombined light
destructively interferes so that the dark port is kept dark.  The GW
signal is read out from the forces needed to keep the recombined light
in destructive interference.}
\label{fig:ligo_optical}
\end{figure}

The LIGO detectors operate as power recycled Michelson interferometers
with Fabry-Perot arms; see Fig.\ {\ref{fig:ligo_optical}}.  Very high
duty cycle is needed for each interferometer in order to effectively
use the full network for coincidence analysis, which is necessary for
achieving a low false detection rate and high confidence
observations.  The wide (3000 km) separation between the LIGO sites is
large enough that the chance of environmentally induced coincidence
events is small.  Both sites are equipped with environmental
sensors that cover a wide range of possible disturbances
that otherwise could
cause false detections.  For example, LIGO monitors the local seismic
background, electromagnetic fluctuations, acoustic noise, cosmic
radiation, dust, vacuum status, weather, power line
transients, and magnetic fields (using ultra-sensitive magnetometers at several
locations at each observatory).

We now briefly describe the operating principles of LIGO, the major
sources of noise that limit sensitivity, and plans for future
upgrades.

\subsubsection {Laser, optics, and configuration}

The basic optical layout of the LIGO detectors is shown in Figure
{\ref{fig:ligo_optical}}.  LIGO uses a Nd:YAG near infrared laser
(wavelength 1064 nm) with peak power $\sim$ 10W as the light source.
Various electro-optical components and servo loops are used to
stabilize both the frequency and power of the laser.

The light from the pre-stabilized laser passes through the input
optics and is coupled into the 12 meter, triangular mode cleaner
cavity.  The mode cleaner passes only the TEM$_{00}$ mode, eliminating
higher order modes.  Starting with the mode cleaner, every major
optical component is within a large vacuum system, operating at
$10^{-9}$ Torr.  After conditioning by the mode cleaner, the light
enters the interferometer.  All major optics in the interferometer are
suspended on a single steel wire loop, mechanically isolated from the
ground by vibration isolators and controlled by multiple servo loops.
The mirrors are made of fused silica with extremely high mechanical
$Q$ and polished to within $\sim$ 1 nanometer RMS.  They have high
homogeneity, low bulk loss, and multi-layer coatings with less than 50
ppm scattering loss.  Each mirror is actuated by four precision coils,
each positioned around a permanent magnet glued to the back side of
the test masses.  The coil assembly also features a sensitive shadow
sensor for local control.  Additional optical levers and wavefront
sensors provide more precise sensing.  The laser beam is coupled into
the arms by a beam splitter.  Each arm is a Fabry-Perot optical
cavity, increasing the effective length of the arm to magnify the
phase shift (proportional to cavity finesse) caused by the wave.  The
stored power within the interferometer is built up by the partially
transmitting recycling mirror.

An operating interferometer tries to keep the dark port perfectly
dark, adjusting the positions of the various optical components such that light coming
out of the arms destructively interferes and no light goes to the
photodetector.  When this is achieved, the interferometer is on {\it
resonance}, with maximum power circulating in the arms.  Several
interconnected control loops are used to achieve and then maintain
resonance.  An interferometer on resonance is usually described as
{\it locked}.  Keeping lock must be highly automated, requiring
minimal operator interaction for high uptime.  The GW signal is
extracted from the servo signals used to maintain the lock and correct
for the changing length difference between the arms.

\subsubsection {Noise sources}

The sensitivity of GW interferometers is limited by a large number of noise
sources.  We list here some of the most important and interesting
fundamental noise sources; many of these were originally recognized,
and had their magnitudes estimated, by Weiss (1972).

\begin{itemize}

\item \textbf{Seismic noise}.  Ambient seismic waves (e.g.,
loading of the continental shelf by surf 
striking the coast) or culturally induced seismic waves
(e.g., passing trucks, logging, cattle guards) continuously
pass under the test masses of the detector. The natural
motion of the surface peaks around 150 mHz; this is called
the ``microseismic peak''.  Cultural noises tend to be at
higher frequencies, near several Hz.  The test masses must
be carefully isolated from the ground to effectively
mitigate seismic noise.  Seismic noise will limit the low
frequency sensitivity of first generation ground-based GW
detectors; the only way to get good performance below
$\sim1\,\mbox{Hz}$ is to put one's detector space. This is a
major motivator for the LISA\footnote{The Laser Interferometer
Space Antenna~(\cite{lisa1,lisa2}).} gravitational-wave detector.

\item \textbf{Thermal noise}.  Thermally excited vibrational modes of
the test mass or the suspension system will couple with the
resonances of the system.  By improving the $Q$ of the
components one can isolate the thermally induced noise to
the resonant frequency.

\item \textbf{Shot noise}.  The number of photons in the input laser
beam fluctuates; this surfaces as noise at the dark port.  This noise
is proportional to $1/\sqrt{{\rm recycling\ gain} \times{\rm input\
laser\ power}}$.  Increasing the recycling gain and/or increasing the
laser power lowers the shot noise.  Unfortunately, high power in the
cavities induces other unwanted effects, such as radiation pressure
noise (discussed in the next item) or thermal lensing (local
deformation of the optical surfaces of the cavity).  The right choice
of laser power and recycling gain involves a compromise
among many sources of noise.

\item \textbf{Radiation pressure noise}.  Fluctuations in the number
of photons reflecting from the mirrors induces a fluctuating force on
the mirrors.  This effect scales as $\sqrt{{\rm recycling\
gain}\times{\rm input\ laser\ power}}$---the inverse of the
proportionality entering the shot noise: there is a penalty
to increasing the laser power.  Reducing shot noise and radiation
pressure noise in tandem is a topic of advanced detector R\&D; see
Buonanno \& Chen (2001a,b) and references therein for further
discussion.

\item \textbf{Gravity gradient noise}.  When seismic waves,
atmospheric pressure fluctuations, cars, animals, tumbleweeds, etc.,
pass near a GW detector, they act as density perturbations on the
neighboring region.  This in turn can produce significant fluctuating
gravitational forces on the interferometer's test masses
(\cite{ht1998,teviet2000}), which is expected to become the limiting
noise source at low frequencies for advanced ground-based detectors
with high quality seismic isolation systems.

\item \textbf{Laser intensity and frequency noise}. All
lasers have some inherent noise, which causes fluctuations
in both the laser's intensity and frequency.  This noise does not cancel
out perfectly when the signal from the two arms
destructively recombines, and so some noise leaks into the
dark port.

\item \textbf{Scattered light}.  Some laser light can scatter out of
the main beam, and then be scattered back, coupling into the
interferometer's signal.  This light will carry information about its
scattering surface, and will generally be out of phase with the beam,
contaminating the desired signal.  A dense baffling system has been
installed to greatly reduce this source of noise.

\item \textbf{Residual gas}.  Any vacuum system contains some trace
amount of gas that is extremely difficult to reduce; in LIGO, these
traces (mostly hydrogen) are at roughly $10^{-9}$ Torr.  Density
fluctuations from these traces in the beam path will induce index of
refraction fluctuations in the arms.  Residual gas particles bouncing
off the mirrors can also increase the displacement noise.

\item \textbf{Beam jitter}.  Jitter in the optics will cause the beam
position and angle to fluctuate slightly, causing noise at the
dark port.

\item \textbf{Electric fields}.  Fluctuations in the electric field
around the test masses can couple into the interferometric signal via
interaction between the field and the induced or parasitic surface
charge on the mirror surface.

\item \textbf{Magnetic fields}.  Fluctuations in the local magnetic
field can affect the test masses when interacting with the actuator
magnets bonded to the surface of the mirrors.

\item \textbf{Cosmic showers}.  High energy penetrating muons can be
stopped by the test masses and induce a random transient due
to recoil.

\end{itemize}

The initial LIGO detectors will be limited by seismic noise at low
frequencies ($\lesssim 50\,{\rm Hz}$), by thermal noise in the test
mass suspensions at intermediate frequencies ($\sim 50$--$200\,{\rm
Hz}$), and by shot noise at high frequencies ($\gtrsim 200\,{\rm
Hz}$).  Present detector noise is above the target level,
particularly at low and intermediate frequencies, though there has
been much progress recently in approaching the target noise curve.

\subsubsection{Detector upgrades}

Much research within the experimental GW community focuses
on developing technologies for improving the sensitivity of
LIGO and other ground-based detectors.  The first stage
detectors are somewhat conservatively designed, ensuring
that they can be operated without the excessive introduction
of new technology.  The price for this conservatism is
limited astrophysical reach: current LIGO sensitivity is
such that detection of sources is plausible, but not
particularly probable, based on our current understanding of
sources.

To broaden the astrophysical reach of these instruments, major
upgrades are planned for 2006/7.  The goal of these upgrades is to
push the lower frequency ``wall'' to lower frequencies, and push the
noise level down by a factor $\sim 10$ across the band.  This will
increase the distance to which sources can be detected by a factor of
$10$ -- $15$, and the volume of the universe which LIGO samples by a
factor of $1000\mbox{--}3000$.  This will dramatically boost any
measured event rates.

Discussion of these plans is given in Gustafson et al.\ (1999).  Major
changes include a redesigned seismic isolation system (pushing the
wall down to about 10 Hz), a more powerful laser (pushing the shot
noise down by about a factor of 10), and replacement of the optical
and suspension components with improved materials to reduce the impact
of thermal noise.  In addition, the system will allow ``tunable''
noise curves---experimenters will be able to shape the noise curve
to ``chase'' particularly interesting sources.

\section{GW emission mechanisms}

Several distinct physical mechanisms could drive GW emission in stellar
core collapse.  Following our earlier discussion (cf.\ Sec.\ 2 of
FHH), we begin with a quick overview of GW generation; further
detail and references can be found in Thorne (1987).  We then discuss
numerical methods of calculating wavestrains given a distribution of
masses and mass currents, and the GWs produced by certain
instabilities---bar modes, fragmentations, and r-modes.

Unlike FHH, we do {\it not} discuss the ringing of newly born black
holes.  A black hole that is distorted from its stationary Kerr
configuration will radiate GWs that drive it back to the stationary
state.  A black hole formed in core collapse will certainly be
distorted; if large amounts of material accrete onto it, it will
be continually driven into new states of distortion.  This in
principle could form an interesting GW source.  In practice,
however, FHH show
that it is not likely to be interesting as a LIGO source: the waves are
emitted at high frequencies where detector sensitivities are
poor, and the associated strain is unlikely to be interestingly
large.  The interested reader should consult FHH for further details.

\subsection{Formal overview}

The conventional approach to calculating the GW emission of a given
mass distribution is via a multipole expansion of the perturbation
$h_{\mu\nu}$ to a background spacetime $g_{\mu\nu}^{\rm B}$.  The
transverse-traceless projection of this metric, evaluated in the
radiation zone, is the metric of the radiation field.  The lowest
(quadrupole) order piece of this field is (\cite{thorne1980})
\begin{equation}
h_{jk}^{\rm TT} = {1\over d} \left[{ 2{G\over c^4} {d^2\over
dt^2}{\cal I}_{jk}(t-r) +{8\over3}{G\over c^5}\epsilon_{pq(j}{d^2\over
dt^2} {\cal S}_{k)p} (t-r)n_q}\right]^{\rm TT}.
\label{eq:thorne1}
\end{equation}
${\cal I}_{jk}$ and ${\cal S}_{jk}$ are the mass and current
quadrupole moments of the source, $d$ is the distance from the source
to the point of measurement, $\epsilon_{ijk}$ is the antisymmetric
tensor, and $n_q$ is the unit vector pointing in the propagation
direction.  Parentheses in the subscripts indicate symmetrization over
the enclosed indices, and the superscript {\rm TT} indicates that one
is to take the transverse-traceless projection.  Most GW estimates are
based on Eq.\ (\ref{eq:thorne1}).  When bulk mass motions dominate a
source's dynamics, the first term describes the radiation that is
generated; for example, it produces the well-known ``chirp''
associated with binary inspiral.  The second term dominates for a
system whose dynamics are dominated by mass currents, as is the case
for radiation from the r-mode instability.

When the background spacetime is flat (or nearly so) the mass and
current moments have particularly simple forms.  For example, in
Cartesian coordinates the mass quadrupole is given by
\begin{equation}
{\cal I}_{jk} =
\int d^3x\,\rho\left[{x^jx^k-{1\over3}r^2\delta_{jk}}\right],
\end{equation}
where $\rho$ is the mass density, and $\delta_{jk}=1$ for $j=k$ and
$0$ otherwise.  The $\delta_{jk}$ term ensures that the resulting
tensor is trace free.

GWs carry energy and angular momentum from the source
(\cite{isaacson}).  The lowest order contribution to the power, $P$,
emitted in GWs is due to variations in the quadrupole moment:
\begin{equation}
P={dE\over dt}={1\over5}{c^5\over G} \left\langle{d^3{{\cal
I}_{jk}}\over dt^3}{d^3{{\cal I}_{jk}}\over dt^3}\right\rangle\;.
\label{quad}
\end{equation}
Although radiated power is useful for understanding the
effect of GW emission on a source's dynamical evolution, $P$
is not useful for detectability estimates.  Instead, one
needs an estimate of the wavestrain $h$, which is directly
measured by the detectors. When averaged over
all source orientations and sky positions, the power and strain are related by
\begin{equation}
P= {\pi^2 c^3\over G}f^2d^2h^2,
\end{equation}
where $f$ is the GW frequency and $d$ is the luminosity distance to
the source.  For a given strain, higher frequency waves radiate more
energy.  Because of detector noise, however, higher frequency waves
are not necessarily more detectable.

\subsection{Bar modes}
\label{subsec:barmodes}

\begin{figure}[t]
\includegraphics[width = 12cm]{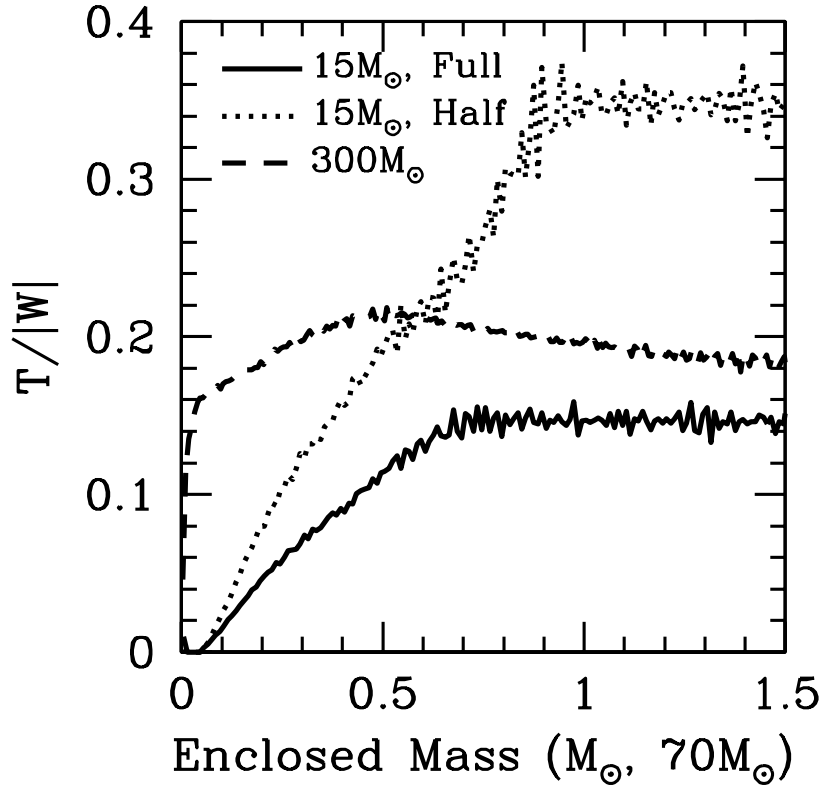}
\caption{Rotational energy divided by gravitational energy
($\beta\equiv T/|W|$) vs.\ mass for collapsing stars.  The horizontal
axis is in solar masses for the $15\,M_\odot$ stellar models, in
fractions of $70\,M_\odot$ for the $300\,M_\odot$ models.  Shown are
rotating core collapse (full rotation), 1.6 seconds after bounce; core
collapse with half rotation, 1.4 seconds after bounce; and direct
collapse of a $300\,M_\odot$ star, 1.9 seconds after bounce.  Note
that $T/|W|$ for the half rotating progenitor is actually larger than
for the fully rotating progenitor.  This is because that star is more
compact.  }
\label{fig:t_over_w}
\end{figure}

Bar modes are an instability in which the material in the
stellar core forms a rapidly rotating bar-like structure.  Such a
structure has a rapidly varying quadrupole moment, and as such is
potentially a copious emitter of GWs.  Bar mode instabilities occur in
objects whose rotational kinetic energy exceeds some fraction of their
potential energy, with the ratio generally written as $\beta\equiv
T/|W|$.  Standard lore (\cite{chandra1969}) tells us that an object is
unstable on a secular time scale if $\beta\gtrsim 0.14$, and is
dynamically unstable if $\beta\gtrsim 0.27$.  This lore may be
violated if the density profile is not centrally peaked: if
centrifugal forces produce a peak in the density off the source's
rotational center, dynamical instabilities can set in at much lower
values of $\beta$ (\cite{cent_etal2000}).  Most core-collapse
simulations with up-to-date progenitor models, however, find that the mass
distribution does not suffer a centrifugal hang-up.  This follows from
the fact that, in these modern progenitor models, the rotation speeds
of the collapsing stellar cores is relatively small
(\cite{heger1998}), particularly when compared to earlier models
(\cite{rampp1998}).  In these simulations, the density is centrally
concentrated, and thus the low critical values of Centrella et
al. (2000) do not apply.  Nonetheless, the rotational energy of these
models can be quite large (cf.\ Fig.\ {\ref{fig:t_over_w}}), so
$\beta$ is likely to be large enough that bar-mode instabilities will
occur.

Heartened by this possibility, we review here expressions for bar mode
GW emission.  Consider a bar of mass $m$ and length $2r$, rotating
with angular frequency $\omega$. The GW energy radiated is given, in
the quadrupole approximation, by
\begin{equation}
P_{\rm bar}={32\over45}{G\over c^5}m^2r^4\omega^6.
\end{equation}
A detector at a distance $d$ from the source would measure an rms
strain
\begin{equation}
h_{\rm bar}=\sqrt{32\over45}{G\over c^4} {m r^2 \omega^2\over d}.
\end{equation}
Note that, due to symmetry, the frequency of the emitted GWs is twice
the bar's rotation frequency.

\subsection{Fragmentation instability}

To set a physically motivated upper limit to the GW emission that
might be produced in stellar collapse, imagine that the collapse
material fragments into clumps, which then orbit for some number of
cycles as the collapse proceeds.  For concreteness, consider the
material fragmenting into a binary system (though it could very well
fragment into more objects).  Such a matter distribution is plausible
if the density distribution during collapse peaks off center, as is
indicated by some simulations of collapsing Population III stars
(\cite{FWH}).

Two bodies, each of mass $m$, in circular orbit about one another at a
frequency $\omega$ and with separation $2r$, radiate GWs with power
and mean strain
\begin{eqnarray}
P_{\rm bin} &=& {128\over5}{G\over c^5}{m^2 r^4\omega^6}\\
h_{\rm bin} &=& \sqrt{128\over5}{G\over c^4}{m r^2\omega^2\over d}.
\end{eqnarray}
These formulae make no assumption about orbital frequency, and thus
apply to, for example, pressure supported as well as Keplerian
orbits. For Keplerian orbits, $4\omega^2 r^3=Gm$, so the above
expressions become
\begin{eqnarray}
P_{\rm bin} &=& {2\over5}{G^4\over c^5}{m^5\over r^5}\\
h_{\rm bin} &=&\sqrt{8\over5}{G^2\over c^4}{m^2\over r\,d}.
\end{eqnarray}
Note that if the ``horizons'' of the two bodies touch ($r=2m\,G/c^2$),
the power radiated reaches a maximum of $P = c^5/80G \sim
10^{57}\mbox{ ergs s}^{-1}$, independent of the system's mass.  The
length of time such emission can be sustained scales with the total
mass---supermassive black hole binaries do radiate more than
microscopic ones.

\subsection{R-modes}

R-modes are peculiar instabilities that may occur in neutron stars.
These modes are large scale oscillations in the {\it current}
distribution of the fluid---they drive very little change in the
star's density, and as such, their GW generation is described using the
second term of Eq.\ (\ref{eq:thorne1}).  R-modes have been of
particular interest to GW studies (e.g.,
{\cite{lom1998,aks1999,olcsva1998}}) because they are unstable to GW
emission: gravitational radiation tends to {\it increase} the
amplitude of the mode.  Lindblom, Owen, \& Morsink (1998) first mapped
out the range of stellar spin and temperature for which the viscosity
was unlikely to damp away this runaway growth, and concluded that many
hot, young neutron stars were likely to be important sources of GWs.
Current thinking is that the r-modes of young hot neutron stars are
actually unlikely to be important sources of GWs.  Most
analyses assume that the growth of the mode will eventually be
limited by nonlinear hydrodynamic effects.  Arras et al.\ (2002) find,
after a careful analysis of mode-mode coupling effects, that the
saturation amplitude of the modes is far smaller---by four orders of
magnitude---than had been assumed previously.  Their rather forceful
conclusion is that r-modes are completely undetectable.

In FHH we examined the r-mode waves that could be produced by various
collapse scenarios that lead to neutron star formation.  Our
conclusions were broadly in line with those of Owen et al.\ (1998).
In fact, we reinforced those conclusions somewhat by pointing out that
material falling back onto the newly-born neutron star could spin it
up and reawaken an r-mode instability that had previously decayed away.
Re-examining the conclusions in light of the wisdom of Arras et al.\
(2002), we are forced to a rather different outcome---the
r-mode waves are unlikely to ever be detected.

As a consequence of this new understanding, we will end our r-modes
discussion here.  Interested readers can re-examine Sec.\ 2.4 of FHH,
noting that the amplitude factor $\alpha$ should be around
$10^{-5}\mbox{--}10^{-4}$, rather than the $0.1\mbox{--}1$ considered
there.  With this in mind, the results for r-modes from young remnants
of stellar collapse (Secs.\ 3.2.2 and 4.2.2 of FHH) are {\it much}
more pessimistic as far as direct detection is concerned.

\subsection{Direct numerical calculation}

\begin{figure}[t]
\includegraphics[width = 11cm]{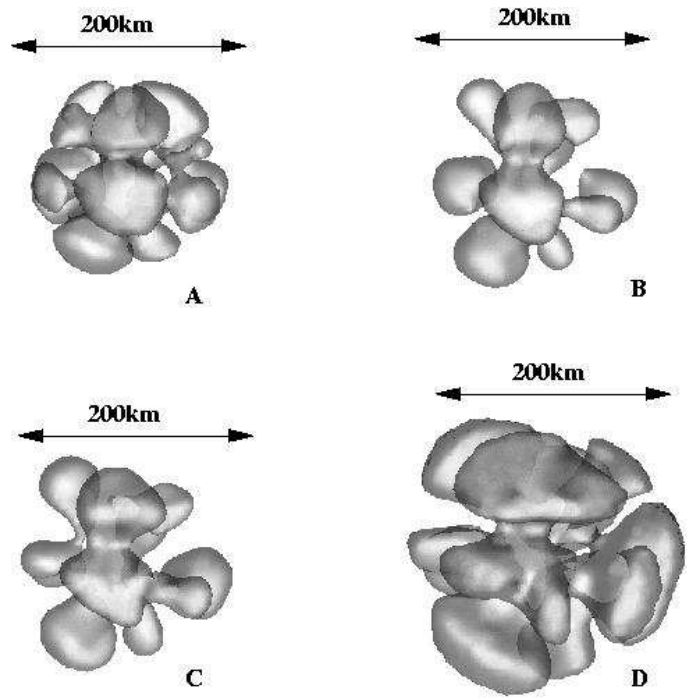}
%\vskip -3 cm
\caption{Isosurfaces of upward moving bubbles (radial velocities
moving outward at 1000 km/sec) as a function of time ($t_A < t_B < t_C
< t_D$) from the 3-dimensional simulations of Fryer \& Warren (2002).
Note that the number of modes decreases slightly as the flows merge
over time (compare panel A with later panels).
}
\label{fig:bubbles}
\end{figure}

If one has a computational model of a core-collapse scenario, one can
calculate the GW emission predicted by that model by numerically
applying Eq.\ (\ref{eq:thorne1}) to the mass and current distributions
predicted by that model.  Naive implementation of Eq.\
(\ref{eq:thorne1}) works poorly: computing the required
derivatives numerically (by evaluating the quadrupole moments on
multiple time slices and differencing), introduces spurious numerical
noise.  However, if the potentials which drive
the motion of the matter are known, the time derivatives can be rewritten
as spatial derivatives, and the GW emission can be calculated using
data from a single time slice.  Calculations demonstrating
this technique are given in Finn \& Evans (1990), Blanchet, Damour, \& Evans
(1990), and Centrella \& McMillan (1993).  The Centrella \& McMillan
paper is specialized to smooth particle hydrodynamics (SPH).

Fryer \& Warren (2002) have recently modeled core collapse with a
3-dimensional code using SPH.  One of the goals of this work will be
to test whether the instabilities discussed above, particularly bar
modes and fragmentation, actually occur in a computational stellar
collapse model.  At present, the only GWs produced by these models
come from the large scale, convective motions of matter near the
protoneutron star~(cf.\ Fig.\ {\ref{fig:bubbles}}).  The GWs from
these motions are computed using the formulae given in Centrella \&
McMillan (1993).  We discuss these results at length below.

\begin{figure}[t]
\includegraphics[width = 11cm]{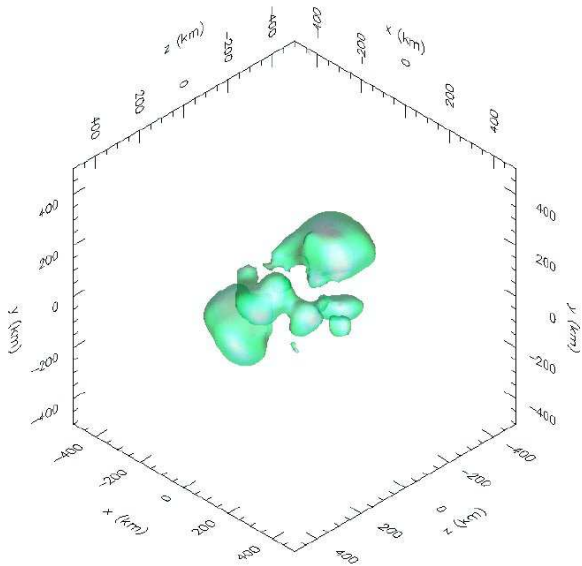}
%\vskip -2 cm
\caption{Isosurfaces of upward moving bubbles (radial velocities
moving outward at 1000 km/sec) from the rotating 3-dimensional 
simulations of Fryer \& Warren (2003).  Due to the angular 
momentum gradient, which prevents convection along the equator, 
most of the convection occurs along the rotation (z) axis.
}
\label{fig:bubble-rot}
\end{figure}

The work of Fryer \& Warren (2002) coupled an equation of state for
dense matter and a flux-limited diffusion neutrino transport scheme
into a parallel 3-dimensional SPH code.  This code includes all of the
physics used in 2-dimensional models less than a decade ago
(e.g. Herant et al. 1994) and shows the remarkable progress in
simulations of core-collapse.  Fig.\ {\ref{fig:bubbles}} shows the
convective upwells of the first such 3-dimensional simulations of the
collapse of a symmetric star.  These simulations assumed spherically
symmetric implementation of gravity for better comparison with past
2-dimensional work.  However, the code developed by Fryer \& Warren
(2002) has a tree-based algorithm for calculating gravity and can be
used to model the collapse of rotating or asymmetric cores.  The first
rotating models are running and, as predicted by the 2-dimensional
simulations (Fryer \& Heger 2000), show convection that is primarily
limited to the polar regions (Fig.\ {\ref{fig:bubble-rot}}).

Improvements still must be made:  a more sophisticated neutrino 
algorithm is needed to accurately model neutrino heating, general 
relativistic and equation of state effects also contribute to 
uncertainties.  However, progress along all these fronts is being 
made.

\section{Results for GW emission}

Applying the various mechanisms discussed in the previous
section, we now estimate GW strengths for two important
stellar collapse scenarios---supernovae, and the death of
Population III stars.  In each case, we discuss what is
known about the rates of these events and their likely
angular momentum distribution (both of which strongly impact
their importance as GW sources), and then estimate their GW
strengths.  In both cases, we sketch the wavestrains likely
for waves arising from bar mode instabilities, and a
possible fragmentation instability, comparing these waves to
the noise level that is the goal of the second generation
LIGO-II detectors (in particular, the broad-band
configuration of those detectors; cf.\ {\cite{whitepaper}}).
For supernovae, we also discuss the recent results of Fryer
\& Warren (2002), for gravitational waves generated in a 3-D
numerical calculation of convective motions in a supernova
explosion.

\subsection{Supernovae}

\subsubsection{Formation rate and angular momentum}
\label{subsec:SNrate}

The supernova rate is fairly well known, lying somewhere between 1 per
50--140 years in the Galaxy (\cite{cappellaro1997}).  It is not clear,
however, what fraction of these core-collapse events (if any) are
rotating rapidly enough to develop these instabilities, and
thereby emit detectable GWs.

Insight into the angular momentum distribution can be gained
by studying pulsars,
the compact remnants of core-collapse supernovae.  From measurements
of young pulsars, we know that at least some neutron stars are born
with periods faster than 20\,ms.  Whether or not any neutron stars are
born with millisecond periods is harder to ascertain---pulsars spin
down as they emit radiation, and the spindown rate is not
particularly well determined. A recent analysis of Chernoff \& Cordes (private
communication) found that the initial spin periods could be fit with a
Gaussian distribution peaking at 7\,ms, with sub-ms pulsars lying
beyond the 2-sigma tail.  Does this mean that less than $10\%$ of
pulsars are born spinning with millisecond periods, or does it mean
that many pulsars are born spinning rapidly, and GW emission removes a
considerable amount of their angular momentum?  It is to be noted that
the analysis of Chernoff \& Cordes is very sensitive to their choice
of spindown rates and other uncertainties in their population study;
they stress that such results should be taken with a great deal of
caution.  Though such large and important uncertainties complicate
efforts to estimate the likely magnitudes of GW emission following
supernovae, they also illustrate the impact that GW observations could
have.  Even a solid null result (no GWs seen from core-collapse with a
high degree of confidence) would have important impact, telling us
that rapidly rotating newly born neutron stars are very rare.

Stellar theorists have now produced models of core-collapse
progenitors which include angular momentum (\cite{heger1998}).  Though
these simulations build in a number of assumptions about the angular
momentum transport in the massive star, they provide some handle on
the angular momentum distribution in the collapsing core.  We base our
analysis on the angular momentum profiles from the core-collapse
simulations of $15\,M_\odot$ stars by Fryer \& Heger (2000), which
use these latest progenitors and model the core-collapse through
supernova explosion (Models 1,5; cf.\ Fig. 2 of FHH).

\subsubsection{GWs from bar modes and fragmentation}

Because the angular momentum distributions used by Fryer \& Heger
(2000) have peak values significantly lower than those used in the
past, there is no centrifugal hang-up.  The collapse proceeds nearly
identically to a non-rotating star, with a density distribution peaked
at the center of the star.  This makes it harder for bar-mode
instabilities to develop, and produces weaker GW emission.  During
bounce, the neutron star is not compact enough to quickly drive
bar-mode instabilities.

However, the explosion produced by these rotating
core-collapse supernovae is much stronger along the poles
than along the equator (\cite{fh2000}), causing much of the
low angular-momentum material to be ejected.  Hence, after
the explosion---$\sim 1$\,s after collapse---$\beta$ can
increase to high enough values that bar-mode instabilities
are likely to develop (cf.\ Fig.\ {\ref{fig:t_over_w}}).
The amount of matter enclosed by the proto-neutron star
extends in all cases beyond $\sim1\,M_\odot$, corresponding
to values of $\beta$ which are certainly above the secular
instability limit ($\beta\sim0.14$), and probably also above the
dynamic instability limit ($\beta\sim0.27$).  Notice in
Fig.~\ref{fig:t_over_w} that $\beta$ is actually larger for the model which
has less initial angular momentum.  This is because this
model has contracted more, and is spinning more rapidly.  
Notice that we have to push to these conditions merely to 
produce bar-mode instabilities.  It is even less likely that 
fragmentation will occur in core-collapse supernovae.

\begin{figure}[t]
\includegraphics[width = 11cm]{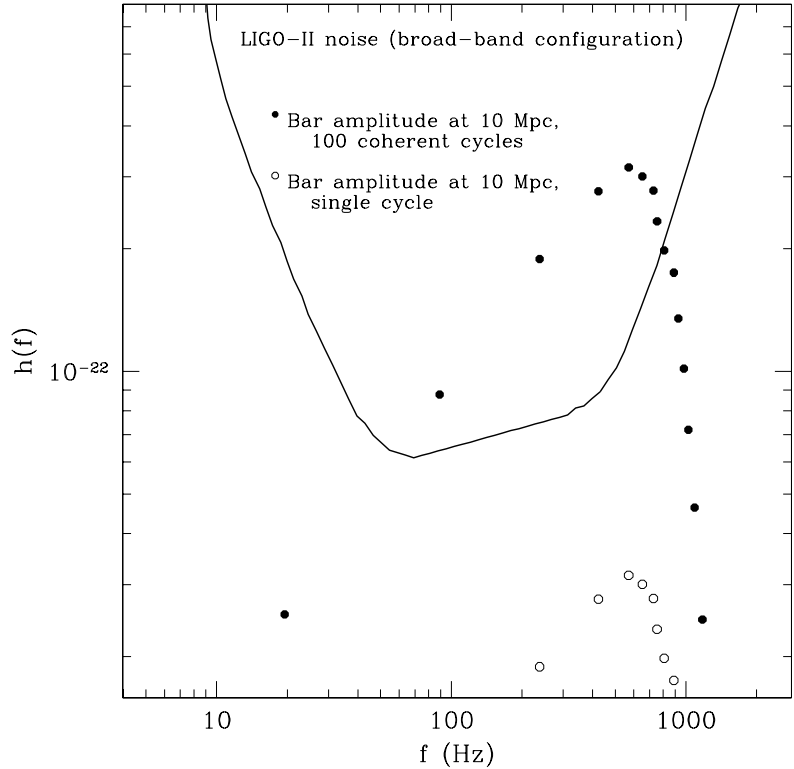}
\caption{Bar mode gravitational waves compared to broad-band LIGO-II
noise.  We use the Fryer \& Heger (2000) models for core collapse, and
then estimate GW emission as follows: we assume that all mass inside a
given radius participates in a bar mode instability and forms a bar,
conserving angular momentum as it does so.  Each point represents the
waves produced by a particular choice of radius, moving to larger
radii from right to left.  An open circle is the strain from a single
wave cycle of the bar; a closed circle is the integrated strain that
would be measured if the bar were to remain coherent for 100 GW
cycles.  The range between the open and closed circles suggests that
bar-mode waves could be of interesting strength provided
that they remain
coherent for a minimum of $\sim 50$--$100$ cycles.
}
\label{fig:bar_15}
\end{figure}

The Fryer \& Heger (2000) simulations are axisymmetric (2D) and so, by
construction, cannot produce a bar mode instability.  However, we can
get a handle on the {\it potential} of these models to radiate by a
bar mode instability by taking the mass and angular momentum
distributions of the Fryer \& Heger models and imagining that some
fraction of the mass in these models participates in such an
instability.  We do this by assuming that all of the matter up to some
enclosed mass becomes unstable and forms a bar (conserving angular
momentum), and then calculate the GW emission as a function of total
unstable mass.  One should bear in mind that we assume {\it all} of
the enclosed mass ends up in the bar.  These estimates are thus
relatively strong upper limits (although the strain could increase if
we allowed the bar to contract and spin up).  We illustrate the
detectability of waves from this bar model in Figure
{\ref{fig:bar_15}}.  Each point on this plot illustrates a different
possible bar, varying the amount of mass that participates in the
instability.  Open circles illustrate the wavestrain for a single GW
cycle; filled circles give the characteristic strain obtainable if the
bar radiates coherently for 100 cycles.  This plot demonstrates that
bar modes are potentially promising sources of waves if bars remain
coherent for at least a moderate ($\sim 50$--$100$) number of GW
cycles.

Since the density is centrally peaked, a fragmentation instability is
unlikely to occur in core-collapse supernovae.  However, if it did
occur, even the single-cycle strain would be quite large ($\sim 10^{-22}$),
and fall in an interesting band (from a few hundred Hz up to a kilohertz or
so).

\subsubsection{Numerically calculated GW estimates}

As computational models improve we can expect to reach the point
where a 3-dimensional numerical calculation can adequately model an asymmetric
supernova explosion, and diagnose whether the various instabilities
discussed above are likely to actually occur.  We're not quite there
yet, but recent progress has been impressive.  As our physical
understanding of supernovae improves, and as more powerful
computational resources become available, we can look forward to serious
numerical studies that model realistic supernova explosions.

\begin{figure}[t]
\includegraphics[width = 11cm]{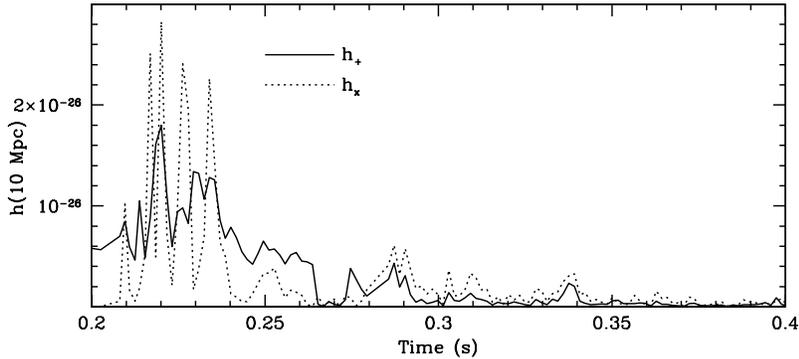}
\caption{Magnitude of the wavestrain for convective motions in a
non-rotating 3-D stellar collapse model (cf.\ Fig.\
{\ref{fig:bubbles}}), for a supernova at 10 Mpc.  }
\label{fig:3dwavestrain}
\end{figure}

At present the most relevant models that can be studied to estimate GW
emission in a 3-D calculation are those produced by Fryer \& Warren
(2002).  The only dynamical aspect of this model with sufficient
asymmetry to produce GWs are the convective cells bubbling through the
stellar core (cf.\ Fig.\ {\ref{fig:bubbles}}; the models are currently
non rotating).  The GW strengths in these bubbles is simply estimated
by postprocessing the model's data using the formulae of Centrella \&
McMillan (1993).  The magnitude of the GW strain expected from this
model's convective behavior is shown in Fig.\ {\ref{fig:3dwavestrain}}
for an explosion at 10 Mpc.

The first rotating simulations are now being run (Fryer \& Warren
2003), but with the rotation speeds expected from stellar evolution
models (\cite{heger1998}), it does not appear that bar-mode
instabilities develop both because of the low, but realistic angular
momentum in the Heger models and because the convection and viscosity
(dominated by numerical viscosity) transport out this angular
momentum.  If bar-modes do not develop, the dominant gravitational
wave signal occurs at bounce where models by Dimmelmeier, Font, \&
M\"uller (2002a,2002b) are among the most physically accurate.

These convective GWs are extremely weak---even though they arise from
the rapid overturning motions of dense matter in the star, the
multipolar distribution of this matter is not ideally distributed to
be a strong source of GWs.  Even in a galactic explosion (distance to
source $\sim 10$ kpc), these waves would be weak ($h\sim10^{-23}$).
Preliminary results from rotating simulations from Fryer \& Warren
(2003) also do not develop strong bar modes and hence do not have
strong GW signals (only a $\sim 2$ order of magnitude increase).  Even
higher angular momenta are required to cause fragmentation.  If the
results of Fryer \& Warren (2003) are correct, fragmentation can be
ruled out in core-collapse supernovae.  Future 3-D models will study
the effects of angular momentum transport and will also include the
effects of asymmetries in collapse to find the maximum gravitational
wave signals that may develop from core-collapse supernovae.

\subsection{Collapse of very massive stars}

At solar metallicity, stellar winds severely limit the pre-collapse
mass of massive stars---very few massive stars will remain massive
up to the time of collapse.  As these winds are driven by the opacity of
metals in the stellar envelope, it is likely that mass loss from
winds will decrease as the fraction of metals in the envelope is
reduced.  Population III stars are the first generation of stars
formed in the early universe, before stars formed the metals that
abound today.  Here we review the death of very massive
($100$--$500\,M_\odot$) Population III stars.  Like Chandrasekhar-massed
white dwarfs, these stars must suffer one of two fates: either they
explode in a giant thermonuclear explosion (``hypernova''), or they
collapse to form black holes.  The fate is determined by the stellar
mass.  If the star's mass exceeds $\sim260\,M_\odot$, it will collapse
to a black hole (\cite{FWH,bhw2001}).  However, if the star is
rotating, rotational (plus thermal) support prevents the star from
undergoing immediate collapse to a black hole (\cite{FWH}). Rotating,
very massive stars collapse and bounce, forming a much larger compact
core than those produced by core-collapse supernovae: a
$50$--$70\,M_\odot$, 1000--2000\,km proto-black hole instead of the
$1\,M_\odot$, 100\,km proto-neutron star.  This rotating proto-black
hole is susceptible to bar instabilities and may produce a strong GW
signal (see also {\cite{mr2001}}).

\subsubsection{Formation rate and angular momentum}

Estimating the rate of core-collapse for very massive stars
depends on two rather uncertain quantities: the amount of matter found
in Population III stars, and the number of these stars which actually
collapse to form black holes.  The mass distribution of stars at birth
is known as the initial mass function (IMF).  Today the IMF is peaked
toward low mass stars, such that 90\% of stellar core-collapse occurs
in stars between 8 and $\sim20\,M_\odot$, and only 1\% of
core-collapse occurs in stars more massive than $40\,M_\odot$.
However, it has long been believed that the first generation of stars
after the Big Bang tended to be more massive than stars formed today
(e.g., {\cite{silk1983,cr1984}}).  Recent simulations by Abel, Bryan,
\& Norman (2000) suggest that the typical mass of first generation
stars is $\sim 100\,M_\odot$, and it is possible that a majority of
Population III stars had masses in excess of $100\,M_\odot$.

The light from these very massive stars re-ionizes the early universe;
from this we can derive a constraint on the formation rate of these
stars.  Although we expect that their photons ionized a significant
fraction of the early universe, there should not be so many stars that
they ionize the universe several times over.  Using our best estimates
of the re-ionization fraction, the amount of ultraviolet photons
produced by these massive stars, and the ionization efficiency of
massive stars, one finds that $0.01\%$--$1\%$ of the baryonic matter in the
universe was incorporated into very massive stars (\cite{abn2000}).
This corresponds to about $10^4$--$10^7$ very massive stars produced in
a $10^{11}\,M_\odot$ galaxy, or a rate of massive stellar collapse as
high as one every few thousand years.

We should temper these optimistic statements with two
caveats.  First, these are Population III stars, and so are
born at high redshift ($z\gtrsim 5$). As they evolve to
collapse in less than a few million years (\cite{bhw2001}),
they will only be observed near to the high redshifts of
their birth. In addition, although we may believe our formation rate
of very massive stars (within a few orders of magnitude), it
is currently impossible to determine what fraction of very massive
stars are produced with masses beyond the $\sim
260\,M_\odot$ mass limit necessary for black hole formation.
The Galaxy could produce a million of these objects, or
perhaps just a few hundred. Assuming 1--10 million very
massive stars per galaxy beyond $z = 5$ gives us a secure
upper limit.

The rotation of these stars has again been calculated using the
stellar evolution code developed by Heger (1998); for this analysis we
use the Fryer et al.\ (2001) rotation profiles.

\subsubsection{GWs from bar modes and fragmentation}

\begin{figure}[t]
\includegraphics[width = 11cm]{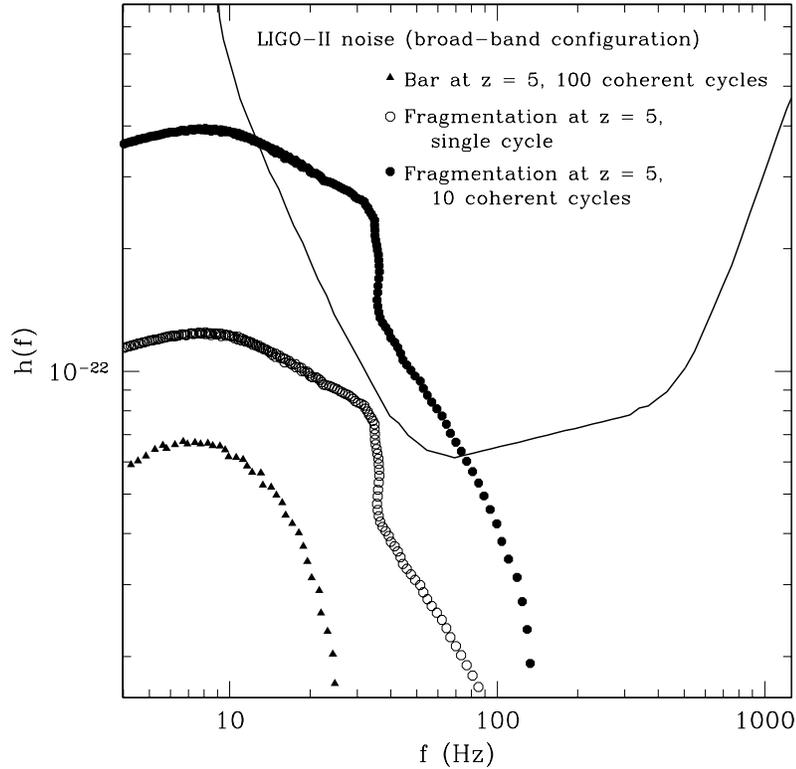}
\caption{Possible bar mode and fragmentation instability GW signals
for collapse of $300\,M_\odot$ Population III star.  As in Fig.\
{\ref{fig:bar_15}}, each point corresponds to all mass inside a
particular radius participating in the instability, conserving angular
momentum.  The prospects for detecting bar modes from these collapse
events are extremely bad: the cosmological redshift pushes this signal
far out of the LIGO band.  Waves generated by a fragmentation
instability are potentially interesting: both their strains and
frequencies are significantly higher, and might be accessible to LIGO,
particularly if the signal remains coherent for some number of
cycles.}
\label{fig:bar_300}
\end{figure}

We expect the proto-black hole formed in the collapse of a massive
star to become secularly unstable (Fig.~\ref{fig:t_over_w}), and these
secular instabilities are likely to develop before the proto-black
hole collapses to a black hole (\cite{FWH}).  Given the
large amount of mass
($\sim70\,M_\odot$) and angular momentum these objects possess, it is
not surprising that these objects can produce strong GW signals.
However, the cosmological redshift moves the peak of the source waves
out of the band of LIGO detectors: even at the relatively low value $z
= 5$ (corresponding to a luminosity distance $\sim48$\,Gpc in the
currently popular ``concordance cosmology'' [\cite{wtz2002}]), the
strain from bar modes peaks at frequencies less than 10 Hz, with a
strain $8 \times 10^{-23}$.  This is well below the LIGO II threshold.
Even coherent integration over $\sim100$ cycles is unlikely to produce
a detectable signal; see Fig.\ {\ref{fig:bar_300}}.

The waves from massive star collapse may be detectable, however, if a
fragmentation instability occurs.  With our crude model of
fragmentation, we find that both strain and frequency are boosted if
the core splits into two pieces which then fall into a Keplerian
orbit, conserving angular momentum.  If this instability occurs and
the pieces orbit coherently for $\sim 10$ cycles, these waves may be
detectable at redshifts $z\sim5$; see Figure~{\ref{fig:bar_300}}.

\section{Summary \& concluding thoughts}

One clear conclusion can be drawn from this survey: despite the
impressive recent progress in our understanding of core collapse, we
{\em still}\/ have a relatively poor grasp of the processes that drive
GW emission in these events.  Past work on instabilities, when coupled
with $\mbox{2-D}$ models of stellar collapse, provide a decent
estimate of the range of wave strengths and frequencies that are
possible, but do not provide specifics for what is to be expected from
collapse.

Progress and future understanding will only come as these models are
further generalized, building in the rotation or core oscillations
that seem needed (\cite{fh2000}) to explain asymmetries in supernovae
(\cite{whhw2001}).  Fully $\mbox{3-D}$ simulations, such as those
from~\cite{fw2002}, hold great promise for extending our knowledge of
the collapse process.  Better stellar models, better neutrino
transport algorithms and the inclusion of general relativistic effects
will allow modelers to unambiguously extract the effects of GWs.  If
the GW detector sensitivities continue to dramatically improve, one
can hope that GW data will meaningfully impact core-collapse work
later in the decade.  As noted in Sec.\ {\ref{subsec:SNrate}}, even a
solid null result would provide useful information.  An unambiguous
{\it non}-null result (perhaps in coordination with other astronomical
instruments) would be most exciting of all, and may open our clearest
window onto the processes occuring deep inside collapsing stellar
cores.

\begin{acknowledgments}
We thank Peter Bender, Craig Hogan, and Szabolcs Mark\'a for allowing
us to use material from Hughes et al.\ (2001), and we thank Kip
Thorne for permission to reproduce Figs.\ {\ref{fig:forcelines}} and
{\ref{fig:ifo}}.  The $\mbox{3-D}$ simulations were funded by a {\it
Scientific Discovery through Advanced Computing} grant, and were
performed on the IBM SP at the National Energy Research Scientific
Computing Center.  The work of DEH and SAH is supported at the KITP by
NSF Grant PHY--9907949.
\end{acknowledgments}

\bibliographystyle{apalike}

\end{document}